# Modelling Animal Biodiversity Using Acoustic Monitoring and Deep Learning

C. Chalmers, P.Fergus, S. Wich and  S. N. Longmore

*Abstrac*t— For centuries researchers have used sound to monitor and study wildlife. Traditionally, conservationists have identified species by ear; however, it is now common to deploy audio recording technology to monitor animal and ecosystem sounds. Animals use sound for communication, mating, navigation and territorial defence. Animal sounds provide valuable information and help conservationists to quantify biodiversity. Acoustic monitoring has grown in popularity due to the availability of diverse sensor types which include camera traps, portable acoustic sensors, passive acoustic sensors, and even smartphones. Passive acoustic sensors are easy to deploy and can be left running for long durations to provide insights on habitat and the sounds made by animals and illegal activity. While this technology brings enormous benefits, the amount of data that is generated makes processing a time-consuming process for conservationists. Consequently, there is interest among conservationists to automatically process acoustic data to help speed up biodiversity assessments. Processing these large data sources and extracting relevant sounds from background noise introduces significant challenges. In this paper we outline an approach for achieving this using state of the art in machine learning to automatically extract features from time-series audio signals and modelling deep learning models to classify different bird species based on the sounds they make. The acquired bird songs are processed using mel-frequency cepstrum (MFC) to extract features which are later classified using a multilayer perceptron (MLP). Our proposed method achieved promising results with 0.74 sensitivity, 0.92 specificity and an accuracy of 0.74.

## I. Introduction

Globally biodiversity is in rapid decline. As a result, there is an urgent need to easily deploy scalable and cost-effective monitoring technology to better model and understand wildlife and the environments they inhabit [1]. Sound is considered to be an important aspect when monitoring wildlife and habitat health. Acoustic sensors provide unobtrusive access to nature, for conservationists and researchers. These sensors provide important ecological data that allows information on abundance, distribution and animal behaviour within ecosystems to be used to model conservation strategies [2]. Typical types of analysis include occupancy or distribution modelling, density estimates and population trend analysis [3]. While camera traps have been the go-to technology in such analysis, acoustic monitoring has been used to extend biodiversity studies. Audio obviously provides a different sensory dimension to images but it also has the added benefit of traversing much larger geographical boundaries and is less impacted by field of sight and the vegetive constraints in many hard to reach environments [4].

Largely due to the geographical reach of acoustic sensors and them being less susceptible to densely populated environments acoustic monitoring is increasing within ecology and conservation and is now considered a key component to understanding animal responses to environmental change [5]. Camera traps have proven to be very useful for detecting large animals. However, when they are combined with passive acoustic monitoring, they can identify a much broader range of animal species that include very small animals not easily detected by camera traps. When acoustic sensors are used in isolation they can be deployed for extended periods (often months) to model a particular ecosystem.

Acoustic sensors generate continuous time-series data and often include a combination of frequencies relating to different signal generators. Different animal species will generate sounds using different acoustic features and frequencies. If is therefore necessary to separate the signal from the noise in order to gain access to required information. Extracting frequency characteristics is most commonly performed using Fast Fourier Transform (FFT). In this paper FFT is implemented in the acoustic monitoring pipeline to generate spectrograms which have previously been used to visually classify and label animal calls [6]. Detection involves locating particular sounds of interest within the recording while assigning each sound to a particular category such as species type. This form of analysis is labour intensive and can often be biased depending on the experience of the conservationist [7]. Figure 1 shows an example spectrogram (House Sparrow) from the dataset used in this paper.

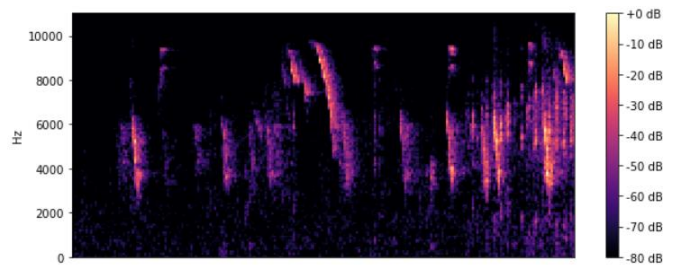

Figure 1. Spectrogram of a House Sparrow

While automated signal analysis has helped to improve classification [8], variability within predictions and efficiency remain a significant issue that impedes widespread adoption [9]. Yet, there is significant interest and support for automated and semi-automated acoustic, including video, analysis among conservationists to speed up study times and facilitate large scale and practical acoustic monitoring.

This paper aims to address these challenges through an automated sound classification pipeline that will help to support large scale acoustic surveys and passive monitoring projects. The current version of the pipeline is capable of classifying different bird songs, although many other types of animals could be included following the generation of species-specific acoustic classification models. Birds have been chosen since they are considered to be an important species when assessing habitat health and modelling biodiversity [10].

The remainder of the paper is structured as follows. A background discussion on current acoustic analysis tools and their associated limitations is introduced in Section 2. Section 3 details the proposed methodology before the results are presented in Section 4. Section 5 discusses the results and the paper is concluded and future work is presented in Section 6.

## II. BACKGROUND AND RELATED WORK

The development of audio classification tools for conservation applications is challenging and often impeded by a number of different factors. These include the availability of validated data, un-biased data (data which is recorded in a variety of different habitats therefore supporting generalisation), standardisation and acoustic tagging [11]. A wide variety of approaches exist and many of them utilise supervised machine learning algorithms such as Artificial Neural Networks (ANNs), Random Forests (RF) and Support Vector Machines (SVMs). The remainder of this section will provide a discussion on some of the more common systems in operation today and highlight their associated limitations that this paper aims to address.

### A. Current Solutions

Historically, the identification of animal species within audio recordings has been undertaken by humans. However, there is now significant interest in fully or semi-automating this process. While, more traditional systems focused on pre-processing audio data to aid in manual classification most approaches now combine pre-processing with automatic classification using machine learning [12]. In existing machine learning approaches, researchers deploy either deep or non-deep learning approaches [13] to classify different animals from acoustic data. These include classifying different animal species such as monkeys, lions and dogs. Studies that primarily focus on same species classifications, such as different birds, have received much less interest amongst machine learning practitioners and conservationists.

This said, a great deal can be learnt from these other more popular studies and their findings mapped directly into within species classification. For example, in [14] researchers developed a convolutional neural network (CNN) to classify different environmental sounds you might find in typical urban settings. The model was evaluated using three different environmental datasets (ESC-50, ESC-10 and UrbanSound8K). While the results reported are relatively low (64.6% accuracy – no sensitivity or specificity values were provided), the paper does provide interesting insights into the development of appropriate pipelines and CNN networks capable of being generalised to animal sounds and acoustic monitoring. This said, a much more in-depth analysis of data pre-processing and network structure is required to improve the results and provide a viable solution in acoustic modelling.

Focusing on animal sounds [15] presents a much more relevant proposition. Again, a CNN architecture is formulated and used to model animal sounds using the Mel Frequency Coefficients (MFCC) library to extract features from audio signals. Unlike the results obtained in [14], [15] was able to able to obtain a classification accuracy of 75%. Again, sensitivity and specificity were not reported.

Directly relating to the approach posited in this paper, several deep learning approaches have been reported in the literature [16] and [17]. In these studies, features extracted from visual spectrogram representations of foreground species recordings were used to train CNNs and achieve 0.605 MAP in BirdCLEF2017. While [10] combined hand-crafted features with deep learning in an attempt to classify fourteen different bird species using three different feature types (acoustic features, visual features, and those generated using deep-learning). They reported that an F1-score equal to 95.95 was possible when all three approaches were combined in an ensemble configuration.

### B. Limitations

CNN approaches require a large corpus of high-quality annotated data that can be used to train the network. Given that there is limited availability of publicly available data that satisfy this requirement there are currently no viable models capable of classifying within species animal types. Another major challenge to overcome is the deployment and automated inference of acoustic sensors. Individually, sensors may generate reasonable amounts of data, but collectively the amount of data that needs to be processed will increase exponentially based on the number of sensors deployed. The first challenge relates directly to how the data is obtained. The second is the cost of compute needed to process the data. Deploying trained models on edge devices for real-time inferencing will take some consideration which has not been sufficiently reported in the literature. Centralising inference will require communications in the field using for example, Global System for Mobile Communications (GSM). The difficulty however is that many environments in which habitat and animal surveys are conducted will not have access to GSM. Again, this issue has not been sufficiently addressed in the literature. Not addressing these issues makes a viable automated acoustic monitoring system less likely.

A perhaps less obvious limitation in the reported literature is the fact that machine learning training and classification is only performed using foreground species. This approach will likely result in poor generalisation once deployed in real world environments. In order to make acoustic classification viable for conservation, foreground and background noise processing must form part of the machine learning pipeline. In the remainder of this paper, we will discuss these limitations further and provide a first-step approach that shows how they may be resolved or mitigated in future acoustic monitoring platforms.

### III. MATERIALS AND METHODS

In this section the dataset used in the study is presented along with the modelling approach taken and the evaluation metrics used to evaluate the trained model. The section also discusses data pre-processing using the Librosa library. Keras and TensorFlow 2.2 are utilised as the backend and an Nvidia 2070 super GPU with 8GB of memory is utilised to accelerate model training. In addition, the proposed inferencing pipeline is discussed along with the associated technologies.

#### A. Data Collection and Description

The audio dataset contains five distinct bird species found in the UK (Lesser Spotted Woodpecker, Eurasian Collared Dove, Great Tit, House Sparrow and Common Wood Pigeon) which is accessible via the Xeno-Canto website[1]. In total the dataset contained 2104 individual wav files. The audio file lengths were variable. In order to standardise the inputs, the audio files were trimmed to the first 15 seconds of the recoding. Figure 2 shows the datasets class distributions. There is a slight class imbalance however this is unlikely to affect the overall performance of the model.

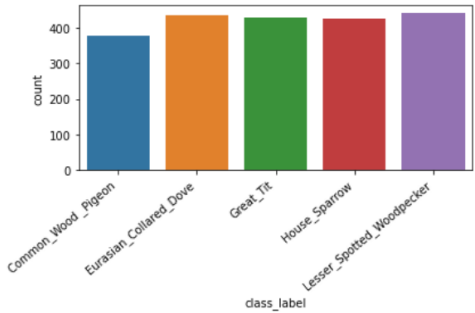

Figure 2. Class Count of Bird Species

Each of the audio files in the dataset were sampled at 44.1kHz. Figure 3 shows an example waveform for each of the classes in the dataset.

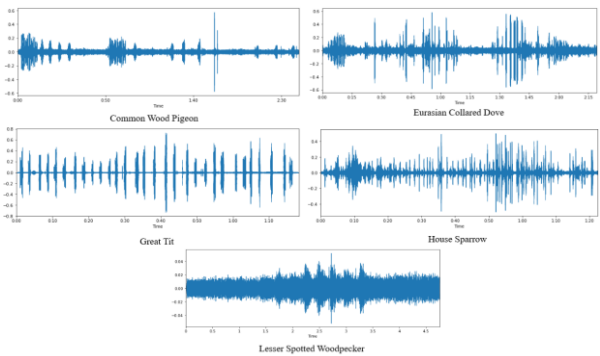

Figure 3. Sample Waveforms

---
[1] https://www.xeno-canto.org/

The dataset contains a limited number of audio files for each of the bird species as shown in figure 2. In addition, the acquired data is comprised of both foreground and background noise of the target class which is reflective of real-world habitats. All of the acquired data is crowd source and requested through the Xeno-Canto website.

#### B. Data Pre-processing

There are a broad range of bit-depths within the dataset (-24440 to 21707) which will to be normalised using the Librosa load function. This is achieved by taking the minimum and maximum amplitude values for a given bit-depth which results in a normalised range between -1 and 1 (-07461247 to 0.66244507). As the dataset contains audio files recorded in both stereo and mono, they are merged to make them uniform. This is achieved by averaging the values of the two channels. Figure 4 shows the original audio file (stereo) at the top and the converted (mono) file at the bottom.

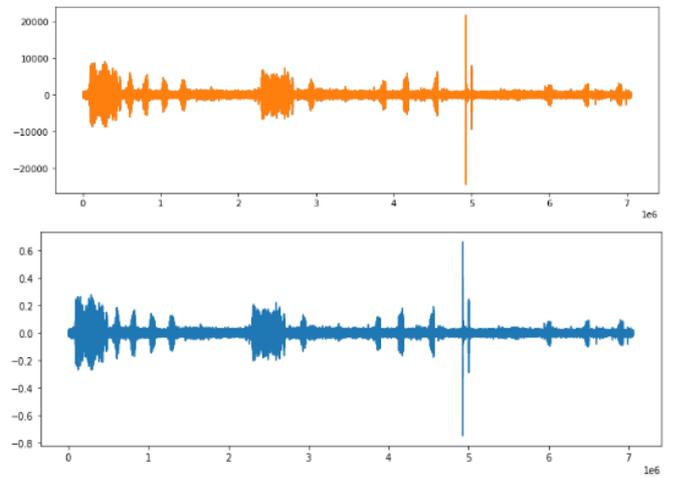

Figure 4. Stereo to Mono Conversion

#### C. Feature Extraction

Features are extracted from the raw audio signals using the Mel-Frequency Cepstral Coefficients (MFCC). MFCC works by summarising the frequency distribution across the specified window size to analyse both the frequency and time characteristics of the acquired audio. The human auditory system does not follow a linear scale. As such for each tone with an actual frequency, f; measured in Hz, a subjective pitch is mapped on a scale called the Mel scale [18]. The process begins by segmenting the audio samples into a reduced frame size of 40msec. Fast Fourier Transform (FFT) is used to convert the N number of samples from the time domain to the frequency domain which is defined as [18]:

$$y(w) = FFT\,[h(t) * X(t)] = H(w) * X(w) \qquad (1)$$

If X (w), H (w) and Y (w) are the Fourier Transform of X (t), H (t) and Y (t) respectively [18]. Bank filters which separate the input signal into multiple components are used to calculate the weighted sum of the filter components which ensures that the output approximates to the Mel scale. Each filter output is the sum of its filtered spectral components. The mel-frequency scale is defined in the following equation where f is the frequency in Hz. The relation between linear frequency and Mel frequency is described as:

$$F(MEL) = [2596 * \log 10[1 + f]700] \quad (2)$$

Discrete Cosine Transformation (DCT) is then used to convert the log Mel spectrum into the time domain. The MFCC window size is set to 80 to capture a broader variety of frequency and time characteristics. Once the MFCC features are extracted the data set is split (train, test) using a ratio of 90/10.

### D. Machine Learning and Modelling

A multilayer perceptron (MLP) is used for the classification task in this study. The network is constructed using the ReLu activation function. ReLU as defined in [20] is:

$$g(x) = max(0,x) \quad (3)$$

The MLP is configured with a filter size of 2 and is used with Backpropagation as the learning algorithm and Adam as the optimiser. A dropout value of 50% has been used in the first three layers to improve generalisation and reduce overfitting. The first three layers are composed of 256 nodes while the final layer is equal to the number of classes in our dataset. The model summary is shown in figure 5.

```
Model: "sequential_1"
_________________________________________________________________
Layer (type)                 Output Shape              Param #
=================================================================
dense_4 (Dense)              (None, 256)               20736
_________________________________________________________________
activation_4 (Activation)    (None, 256)               0
_________________________________________________________________
dropout_3 (Dropout)          (None, 256)               0
_________________________________________________________________
dense_5 (Dense)              (None, 256)               65792
_________________________________________________________________
activation_5 (Activation)    (None, 256)               0
_________________________________________________________________
dropout_4 (Dropout)          (None, 256)               0
_________________________________________________________________
dense_6 (Dense)              (None, 256)               65792
_________________________________________________________________
activation_6 (Activation)    (None, 256)               0
_________________________________________________________________
dropout_5 (Dropout)          (None, 256)               0
_________________________________________________________________
dense_7 (Dense)              (None, 5)                 1285
_________________________________________________________________
activation_7 (Activation)    (None, 5)                 0
=================================================================
Total params: 153,605
Trainable params: 153,605
Non-trainable params: 0
_________________________________________________________________
Pre-training accuracy: 19.9052%
```

Figure 5. Summary of the Compiled Model

The MLP was trained over 100 epochs as the results show this was a sufficient number for the model to converge without overfitting. This section concludes the methods used in this paper to train the model.

The performance of the trained model is measured using Sensitivity, Specificity, Precision and Accuracy. The Sensitivity describes the true positive rate while the Specificity describes the true negative rate. Precision is used to show the number of correctly classified species.

### E. Model Inferencing

The trained model is hosted using TensorFlow 2.2 and served through a public facing website developed by the authors [2]. CUDA 11 and cuDNN 7.6.5 enables the GPU accelerated learning aspect of the pipeline. A Samsung S10 is used to record garden birds and automatically upload the acquired audio to the platform using the Simple Mail Transfer Protocol (SMTP) for classification. Figure 6 shows the end-to-end inferencing pipeline starting with the sensor and ending with the public facing conservationAI site as shown in Figure 7. Due to the use of standard protocols, the system can interface with a variety of sensors for real-time inference. Where in field communication is unavailable, audio files can be batch uploaded through the website for offline inferencing.

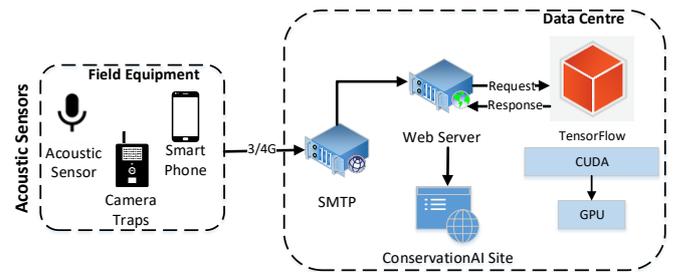

Figure 6. End-to-end Inferencing Pipeline

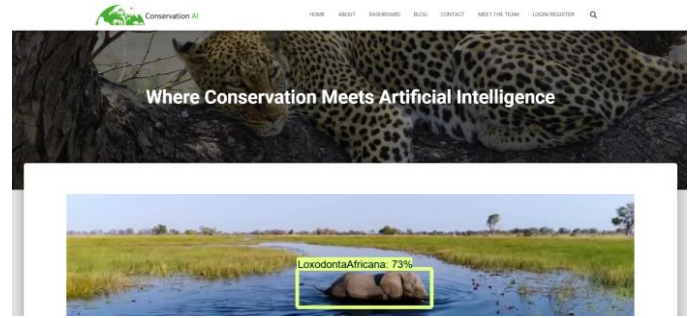

Figure 7. ConservationAI Platform

Inferencing is undertaken on a custom-built server containing an Intel Xeon E5-1630v3 CPU, 64GB of RAM and a NVidia Quadro RTX 8000 GPU. Figure 8 shows the individual stages of the inferencing data pre-processing stages.

---

[2] www.conservationai.co.uk

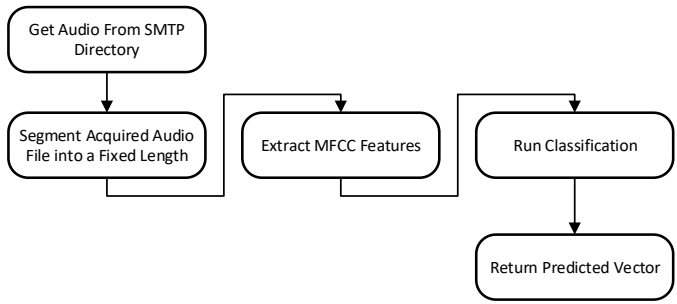

Figure 8. Date Pre-processing Stages

The acquired audio files are transmitted over 4G using SMTP. The audio file is segmented into 15 second windows. Each of the sample windows are passed to the feature extractor function where MFCC is used to return the extracted features for the classifier. The predicted vector is processed and logged to the site for review.

## IV. EVALUATION AND DISCUSSION

In this section the classification results are presented using the evaluation metrics outlined previously. The deployment and inferencing of the trained model in test environment are also presented to ascertain the effectiveness of the end-to-end pipeline.

### A. Species Classification Performance

The results presented in this section were obtained over 100 epochs. Figure 9 shows the loss of the model using both the test and validation data during model training. The figure shows that there was no overfitting during training and that the dropout layers helped with model regularisation. Although model convergence was achieved early in the training session the loss shows continuing decreases throughout the specified epochs.

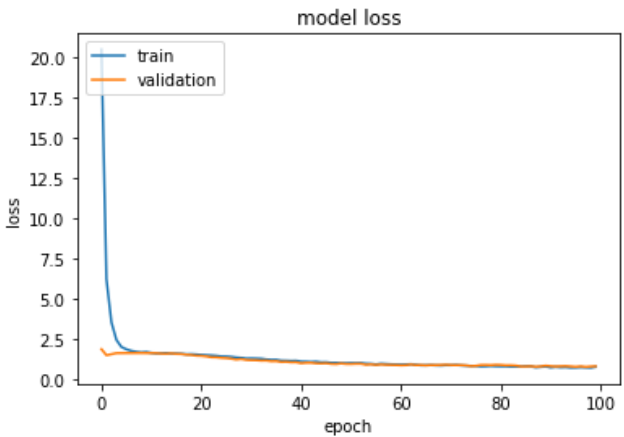

Figure 9. Train and Validation Loss

The model achieved an accuracy of 0.83 for the train split and 0.74 for the test split. Figure 10 shows the accuracy for both the train and validation data over 100 epochs. The results illustrate that the accuracy of the model flattens towards the end of the training session and shows that the necessary number of epochs required for model convergence is sufficient. Increasing the number of epochs would achieve minimal gains in accuracy and would likely lead to overfitting.

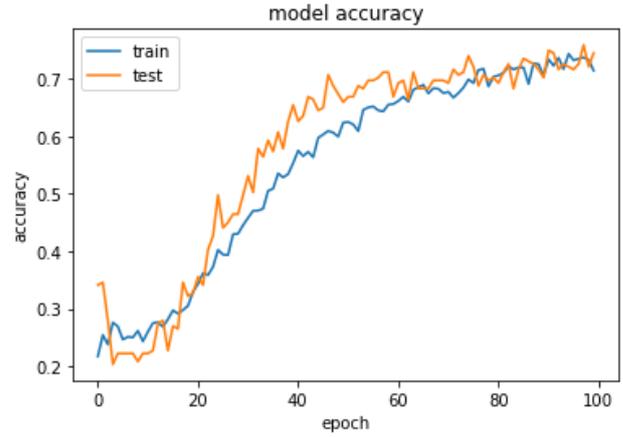

Figure 10. Train and Validation Accuracy During the Training Session.

Table 1 shows the performance metrics obtained using the test data. The best performing class was the Eurasian Collard Dove achieving a Sensitivity of 0.86 and a Specificity of 0.90. The worst performing class was the Lesser Spotted Woodpecker where the model attained a Sensitivity of 0.58 and a Specificity of 0.91.

Table 1. Performance Metrics for Test Set

| Species | Sensitivity | Specificity | F1 Score | Precision |
|---|---|---|---|---|
| Common Wood Pigeon | 0.67 | 0.96 | 0.75 | 0.86 |
| Eurasian Collared Dove | 0.86 | 0.90 | 0.80 | 0.75 |
| Great Tit | 0.91 | 0.92 | 0.82 | 0.74 |
| House Sparrow | 0.75 | 0.91 | 0.72 | 0.70 |
| Lesser Spotted Woodpecker | 0.58 | 0.91 | 0.64 | 0.71 |

### B. Deployment Evaluation

The trained model was implemented in the inferencing pipeline to record bird audio in a realistic environment. This was achieved using a Samsung S10 deployed under a tree containing nesting Common Wood Pigeons. Audio was recorded for a total of three minutes and uploaded to the platform for classification. During deployment 8 individual bird songs were detected. Figure 11 shows an example audio detection on the conservation platform.

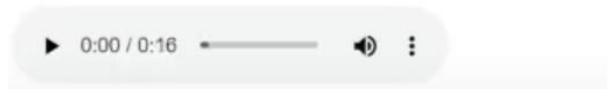

Figure 11. Sample Audio Detection from the ConservationAI Platform

Each of the 8 classifications returned the prediction of Common Wood Pigeon with an average confidence value of 0.71.

## V. DISCUSSION

In this paper we proposed a methodology and pipeline for the classification of five common UK birds – a Common Wood Pigeon, Eurasian Collared Dove, Great Tit, House Sparrow and Lesser Spotted Woodpecker. By extracting features using MFCC and an MLP for classification we were able to achieve encouraging results using a restricted amount of data with limited pre-processing. The results show that bird species can be detected and classified with a reasonable degree of accuracy to rapidly speed up the time taken to manually classify acoustic data. The performance values across all bird classes are encouraging and in many cases are capable of detecting birds with high Sensitively and Specificity values.

There are a number of key advantages of using the proposed methodology. Firstly, there are reduced computational requirements needed to both train and inference a model making it accessible and cost-effective for conservationists. This is in contrast to the approaches reported in the literature as discussed in this paper. While CNNs are used for the classification of bird audio, the data is carefully choreographed to only include foreground noise which represents an unrealistic account of animals in their natural habitat. In our approach we showed that using MFCC we can train the model on more realistic datasets containing both background and foreground noises of the target species. This enables the approach to make use of a wider range of datasets. The initial results are encouraging; however, we envision better performance following the collection of a much larger dataset.

The deployment of the model demonstrates that the system can be used in practical way to automatically classify bird sounds within their natural habitat. Although a Samsung S10 was used in the implementation, a broad range of acoustic sensors could be integrated into the system to achieve the same effect. The inferencing pipeline offers a scalable and cost-effective way to collect, process and classify acoustic audio samples.

## VI. CONCLUSION AND FUTURE WORK

Acoustic data are an import tool to quantify biodiversity and species densities as well as providing assessments of the overall acoustic health of the habitat in which they occupy. Until recently, the processing and classification of the acquired data was largely a manual process therefore limiting widespread deployment. Although advancements have been made in the automatic classification of audio within the conservation domain, significant challenges remain which impede its widespread adoption. The solution presented in this paper overcomes both the computational and dataset limitations outlined in the many existing approaches. This facilitates a scalable and cost-effective solution for automatic acoustic classification.

While a limited range of species have been used in this study, future work will significantly expand the number of classes in the model. In addition, the flexibility of the proposed approach means that it can be rapidly adapted for other scenarios. Such applications include the identification of illegal activity and detecting forest fires using ambient noise or by measuring the stress of species within the affected habitat.

The motivation of this work was to extend our existing pipeline which can already detect animals using vision-based sensors such as camera traps, drones and smart phones as shown in Figure 12.

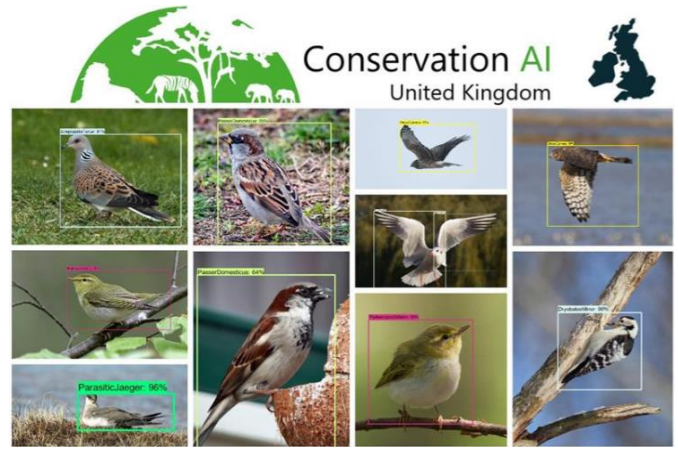

Figure 12. Bird Classification Using Visual Data

By using a combination of both vision and acoustic based data we can extend the reach of the platform into habitats where visual monitoring is not feasible. By studying both image and acoustic data the system can analyse and provide a more holistic overview of the habitat.